\documentclass[a4paper]{jpconf}
\usepackage{graphicx}
\begin{document}
\title{Analysis of anisotropic transverse flow in Pb-Pb collisions at 40$A$~GeV in the NA49 experiment}

\author{Oleg Golosov$^1$, Ilya Selyuzhenkov$^{1,2}$, Evgeny Kashirin$^1$}

\address{$^1$ National Research Nuclear University MEPhI (Moscow Engineering Physics Institute), Kashirskoe highway 31, Moscow, 115409, Russia}
\address{$^2$ GSI Helmholtzzentrum f¨ur Schwerionenforschung, Darmstadt, Germany}

\ead{oleg.golosov@gmail.com}

\begin{abstract} 
Anisotropic transverse flow is one of the most important observables in the study of ultra-relativistic nucleus-nucleus collisions. Detector acceptance non-uniformity in the transverse plane introduces substantial bias in the flow analysis dictating the need for specific corrections. The results of flow analysis in Pb-Pb collisions at the beam energy of 40$A$~GeV recorded with the fixed target experiment NA49 at the CERN SPS are presented. The three-subevent technique is used for the differential measurements of the directed and elliptic flow. Corrections for the detector acceptance anisotropy in the transverse plane are applied using an extension of the Qn-Corrections Framework developed originally for the ALICE experiment at the LHC. The results are compared with those previously published by the STAR at RHIC and the NA49 at CERN SPS collaborations. In the future, the developed procedure will be used for the analysis of the new Pb-Pb data collected by the NA61/SHINE experiment at the CERN SPS.
\end{abstract}

\section{Introduction}
Anisotropies in momentum distributions of particles produced in heavy ion collisions are highly sensitive to the properties of the system very early in its evolution. The origin of this phenomenon lies in the initial asymmetries in the geometry of the system. Anisotropic transverse flow is quantified by the Fourier coefficients $v_n$ of a decomposition of the distribution of particle azimuthal angle $\phi$ relative to that of the reaction plane $\psi_{RP}$~\cite{Selyuzhenkov}:

\begin{equation}
\rho(\phi - \psi_{RP}) = \frac{1}{2\pi} \left( 1 + 2\sum\limits_{n=1}^{\infty}v_{n}(p_T, \eta)\cos \left[ n \left( \phi - \psi_{RP} \right) \right] \right),
\end{equation}

\noindent where $\phi$ is the azimuthal angle of produced particles and $\psi_{RP}$ the azimuthal angle of the reaction plane defined by the beam direction and the impact parameter of the colliding nuclei.

This distribution, obtained in the experiment, can be significantly affected by non-uniformity of detector acceptance in the transverse plane which introduces substantial bias in the flow analysis. 
This effect needs to be corrected using specific methods. In the approach used for this analysis acceptance corrections are determined directly from experimental data~\cite{Selyuzhenkov}.
The aim of this work is to develop a flow reconstruction procedure reproducing previously published results and to use it for the analysis of new data from NA61/SHINE experiment.

\section{Experimental data}
Minimum bias data from Pb+Pb collisions collected by the NA49 experiment at 40$A$~GeV (periods 01D and 02C) were used. Event selection yielded 335K events. Event classification was based on the multiplicity of produced particles registered by the tracking detectors. Particle identification was carried out based on mean energy loss in the VTPC1, VTPC2, MTPC-L and MTPC-R detectors.

\section{Method for flow analysis}
Flow analysis was performed using information from the tracking detectors VTPC1, VTPC2, MTPC-L and MTPC-R. Flow coefficients were calculated with the scalar product three subevent technique described in this section. For each event the produced particles were assigned to three non-overlapping subevents. For each subevent so called flow vectors were calculated, which represent an estimate of the reaction plane orientation\cite{Voloshin} and are defined as:
  
\begin{equation}
  Q_{n} = Q_{x,n} + i Q_{y,n} = \sum u_{n},
\end{equation}

\begin{equation}
  u_{n} = u_x + i u_y = \cos n \phi + i \sin n \phi,
\end{equation}

\noindent where $n$ labels the harmonic. 

Flow vectors were normalized to subevent multiplicity according to the scalar product method~\cite{Selyuzhenkov}. Subevent flow vectors were corrected for detector acceptance non-uniformity using a three-step procedure described in Ref.~\cite{Selyuzhenkov} and implemented in a software QnCorrections Framework originally developed for the ALICE experiment~\cite{FVC}. The values of applied corrections varied depending on event class.

Flow coefficients were calculated as the correlation between particle unit vector $u_n$ and the subevent flow vector $Q_n$ divided by a factor $R_n$ correcting for the finite resolution of the event plane estimation for 
this subevent. The following formulae were used for the case of subevent $a$:

\begin{equation}
v_{i,n}^a (p_T, \eta) = \frac {2 \langle u_{i,n}(p_T, \eta)Q_{i,n}^a\rangle}{R_{i,n}^a},
\end{equation}

\begin{equation}
  R_{i,n}^a\{b,c\} = \sqrt {2 \frac {\langle Q_{i,n}^a Q_{i,n}^b \rangle \langle Q_{i,n}^a Q_{i,n}^c \rangle} {\langle Q_{i,n}^b Q_{i,n}^c \rangle}},
\end{equation}

\noindent where $i = x,y$ indicates the component of the event flow vector chosen for flow calculation based on which detector acceptance non-uniformity effect was smaller for the given harmonic. Analogous formulae were used for subevents $b$ and $c$. An average value of flow was obtained for the three subevents. Errors were calculated using the bootstrapping procedure with 100 subsamples. The definition of $Q_n$ and $u_n$ vectors used in this analysis can be found in Table \ref{Subevents}.

\begin{table}[h]
\caption{\label{Subevents}The definition of the group of particles used to calculate event and unit flow vectors.}
\begin{center}
\begin{tabular}{ccccc}
\br
		&Particle\ species	&$p_T$ [GeV/c]&\multicolumn{2}{c}{Rapidity} \\
\mr
		&		&			&$Q_{1,x}$& $Q_{2,y}$\\
\mr
$Q_n^a$ & $\pi^+$ & 0.1, 2.0	& 1.0, 3.0 & -1.0, 1.0 \\
$Q_n^b$ & $p$	 & 0.1, 2.0	& 1.0, 3.0 & -1.0, 1.0 \\
$Q_n^c$ & $\pi^-$ & 0.1, 2.0	& 1.0, 3.0 & -1.0, 0.0 \\
$u_n$ 	& $\pi^-$ & 0.1, 2.0 & -1.0, 1.0& 0.0, 1.0\\
\br
\end{tabular}
\end{center}
\end{table}

\section{Results}
\begin{figure}[h]
\begin{minipage}{19pc}
\includegraphics[width=19pc]{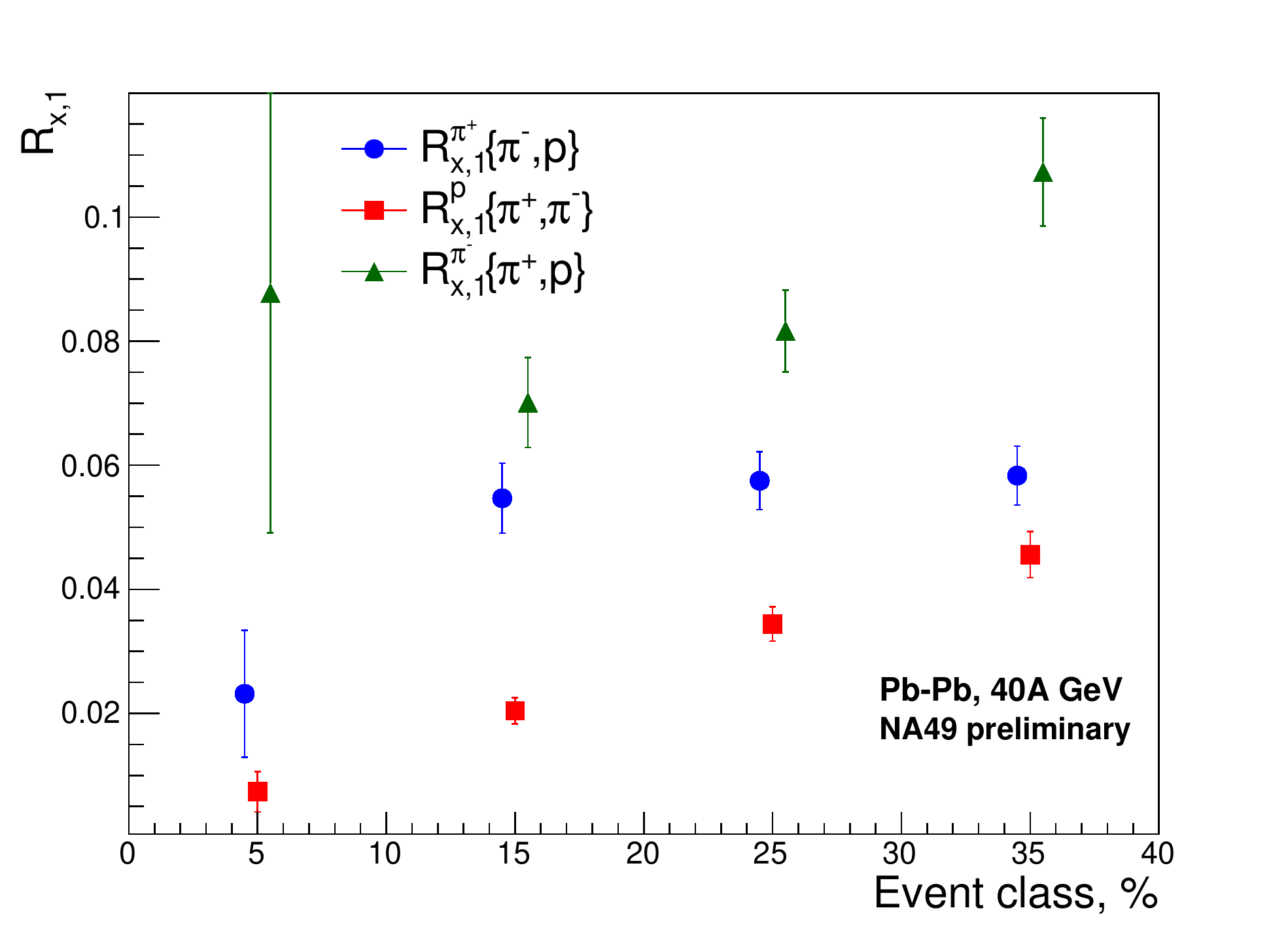}
\end{minipage}
\hspace{0pc}
\begin{minipage}{19pc}
\includegraphics[width=19pc]{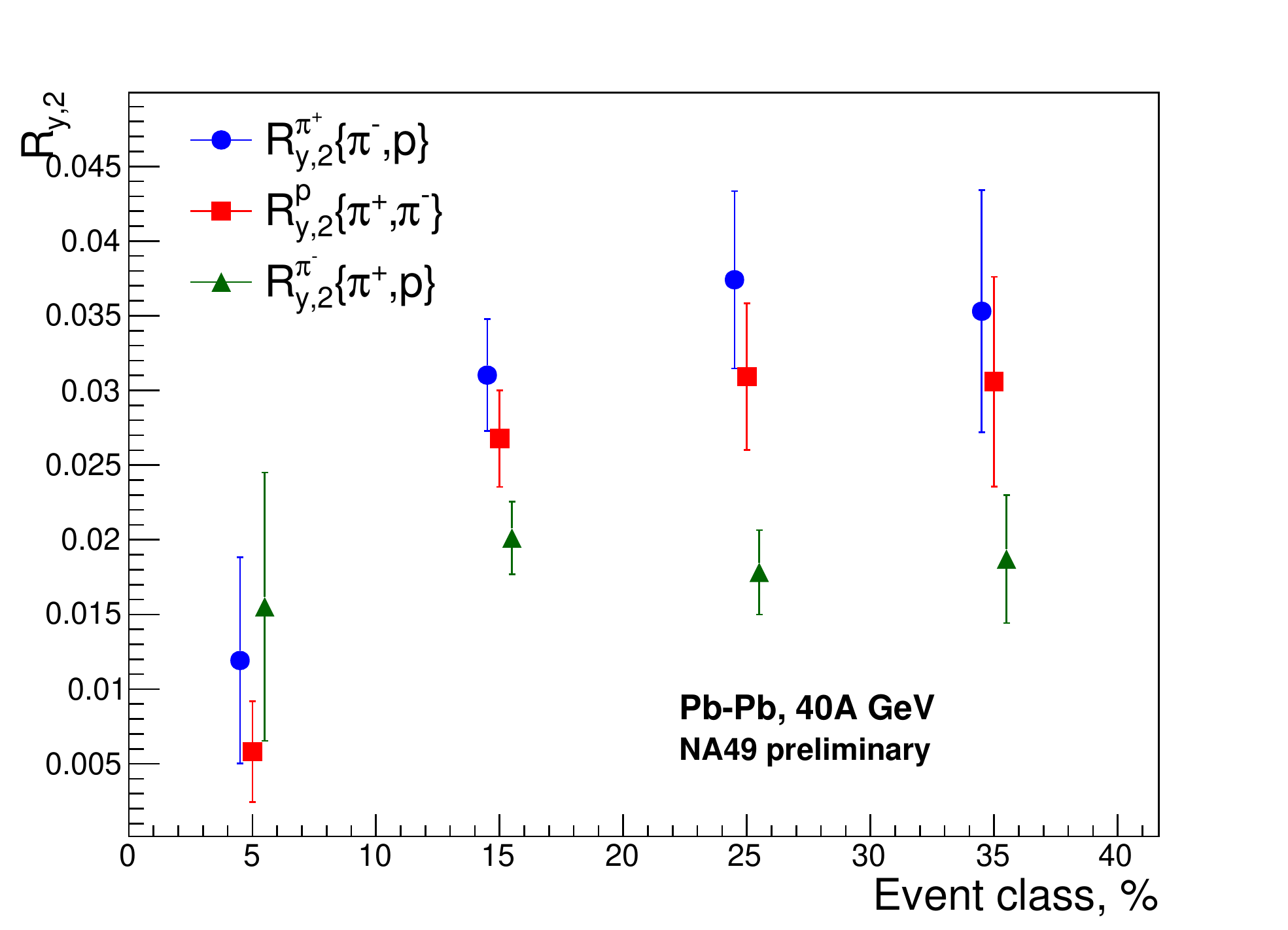}
\end{minipage} 
\caption{\label{Resolution}Resolution correction as a function of event class: $R_{x,1}$ (left) and $R_{y,2}$ (right).}
\end{figure}

The values of resolution correction factors for the three subevents are shown in Fig.~\ref{Resolution} separately for x and y components. A strong dependence on event centrality can be seen in both cases.

\begin{figure}[h]
\begin{minipage}{19pc}
\includegraphics[width=19pc]{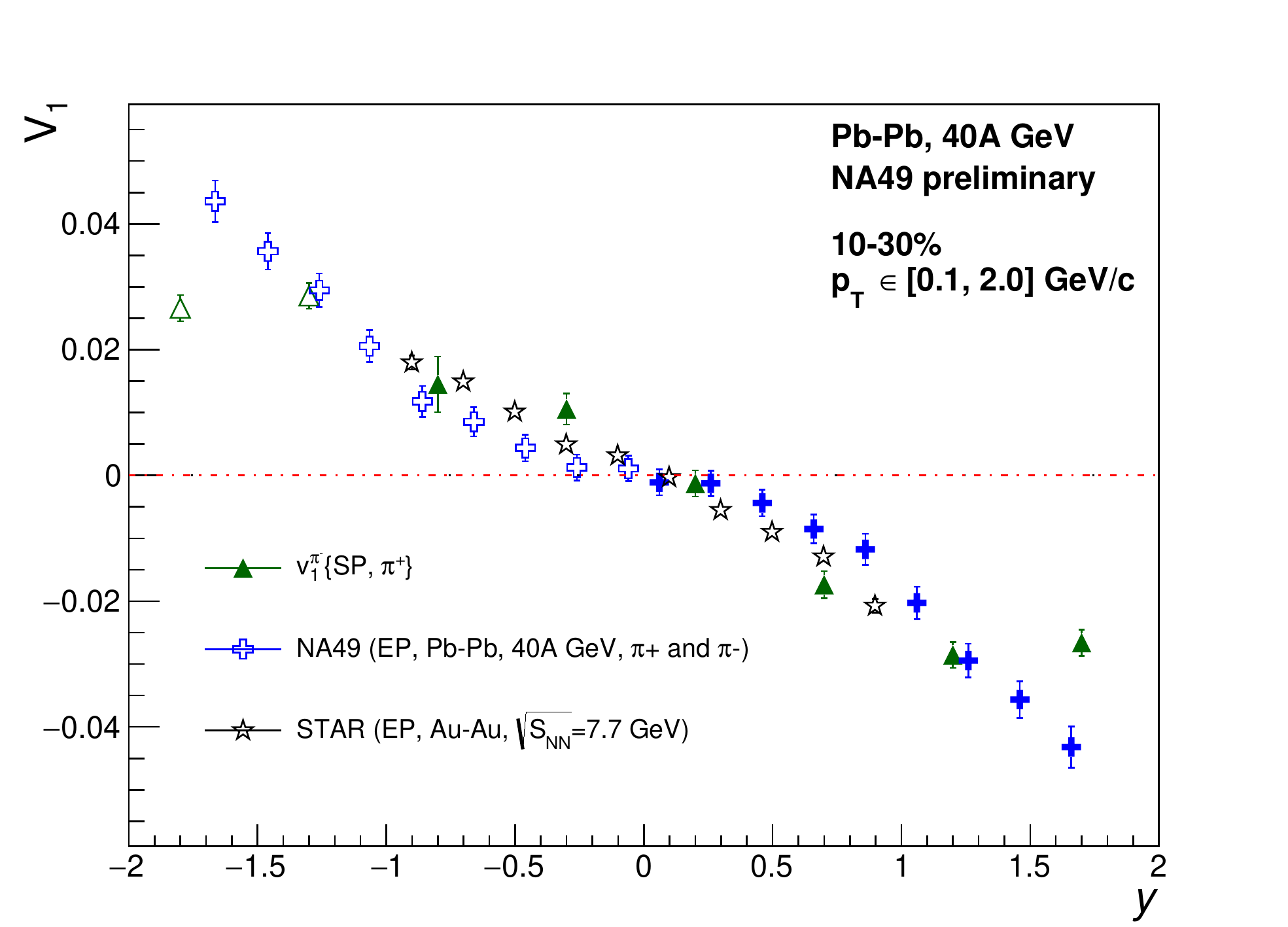}
\end{minipage}\hspace{0pc}%
\begin{minipage}{19pc}
\vspace{1.2pc}
\includegraphics[width=19pc]{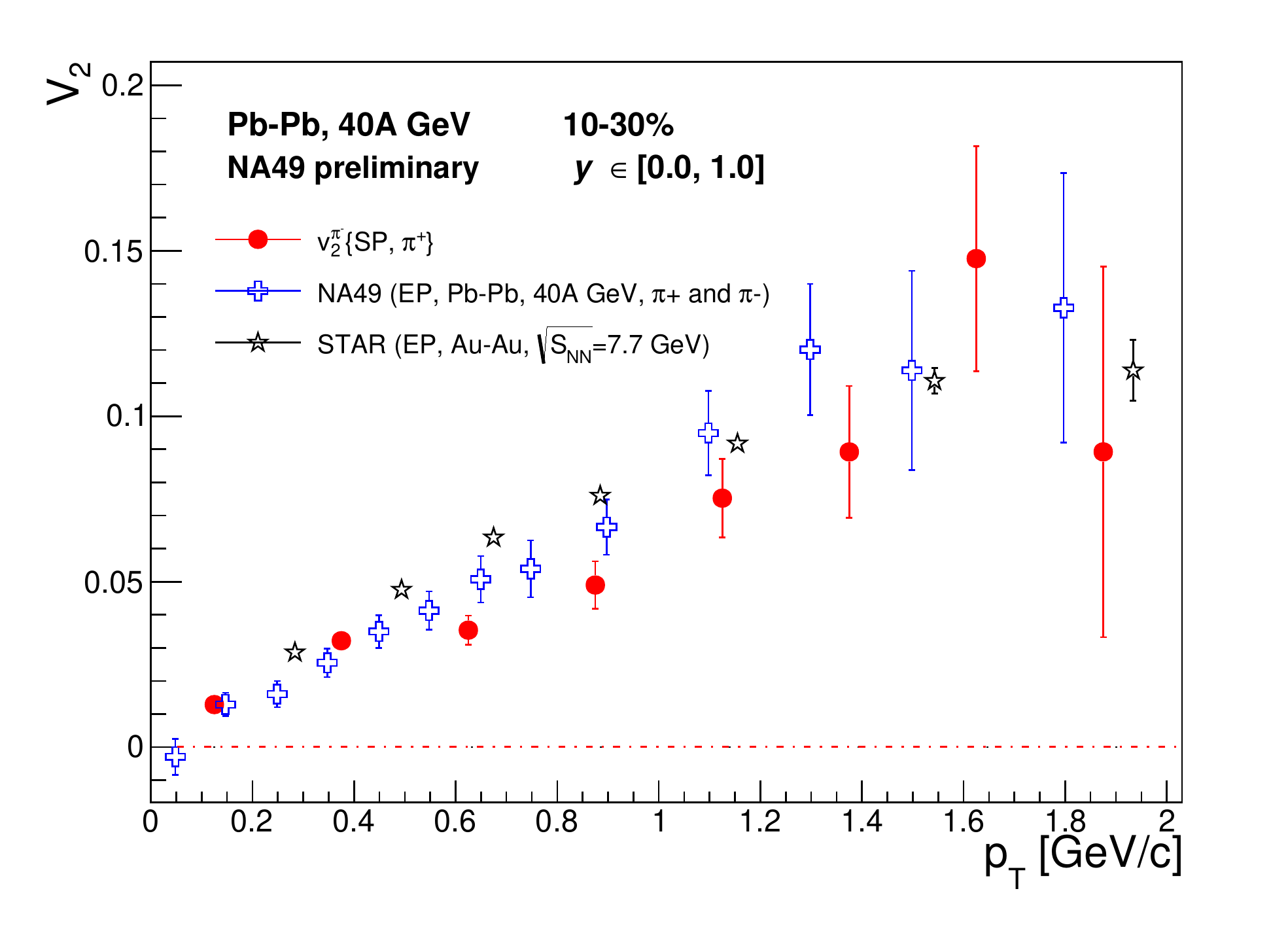}
\end{minipage} 
\caption{\label{V}$v_1(y)$ (left) and $v_2 (p_T)$ (right) for $\pi^-$ (centrality 10-30\%) calculated relative to different subevents. Open points for $v_1(y)$ were reflected at midrapidity. Results are compared with those published by the NA49 (Pb+Pb at 40$A$~GeV, $\pi^+$ and $\pi^-$, random subevent, event plane method) and STAR (Au-Au at $\sqrt{S_{NN}}$ = 7.7 GeV, random subevent, event plane method) collaborations.}
\end{figure}

Results for $v_1(y)$ and $v_2 (p_T)$ for $\pi^-$ in midcentral collisions (centrality 10-30\%) calculated relative to different subevents are shown in the Fig.~\ref{V}. Open points for $v_1(y)$ were reflected anti-symmetrically with respect to zero rapidity. In first approximation $v_1$ appears to be an odd function of rapidity, though it does not cross the zero exactly. This might represent additional correlations introduced by global momentum conservation. Additional corrections should be made to reduce this effect~\cite{momentum}. The results are consistent with those published by the NA49~\cite{NA49} (Pb+Pb at 40$A$~GeV, $\pi^+$ and $\pi^-$, random subevent, event plane method) and STAR~\cite{STAR} (Au+Au at $\sqrt{S_{NN}}$ = 7.7 GeV, random subevent, event plane method) collaborations. 

\section{Summary}
Directed $v_1$ and elliptic $v_2$ flow of $\pi^-$ was measured as a function of transverse momentum and rapidity. The results are qualitatively consistent with those published by the NA49~\cite{NA49} (Pb+Pb at 40$A$~GeV) and STAR~\cite{STAR} (Au+Au at $\sqrt{S_{NN}}$ = 7.7 GeV) collaborations. Further improvement of the method, especially mitigation of effects introduced by global momentum conservation, is under investigation. 

\section*{Acknowledgments}
This work was partially supported by the Ministry of Science and Education of the Russian
Federation, grant N 3.3380.2017/4.6, and by the National Research Nuclear University MEPhI
in the framework of the Russian Academic Excellence Project (contract No. 02.a03.21.0005,
27.08.2013).

\end{document}